\documentclass[journal=jctcce,manuscript=article]{achemso}

\usepackage[version=3]{mhchem} 
\usepackage{nicematrix} 
\usepackage{epstopdf}
\usepackage[utf8]{inputenc}
\usepackage{subcaption}
\usepackage{graphicx}
\usepackage{pdflscape}
\usepackage{pdfpages}

\newcommand\revision[1]{\textcolor{black}{{#1}}}

\newcommand*\rv{\mathbf{r}} 
\newcommand*\rvh{\mathbf{r}_\mathrm{h}} 
\newcommand*\rve{\mathbf{r}_\mathrm{e}} 

\author{Mauricio Rodríguez-Mayorga}
\affiliation[]{Grenoble Alpes University, CNRS, Grenoble INP, Institut Néel, 25 rue des Martyrs, 38042 Grenoble, France}
\author{Xavier Blase}
\affiliation[]{Grenoble Alpes University, CNRS, Grenoble INP, Institut Néel, 25 rue des Martyrs, 38042 Grenoble, France}
\author{\\Ivan Duchemin}
\affiliation{Grenoble Alpes University, CEA, IRIG-MEM-L Sim, 38054 Grenoble, France}
\email{ivan.duchemin@cea.fr}
\author{Gabriele D'Avino}
\affiliation[]{Grenoble Alpes University, CNRS, Grenoble INP, Institut Néel, 25 rue des Martyrs, 38042 Grenoble, France}
\email{gabriele.davino@neel.cnrs.fr}

\title{From many-body \emph{ab~initio} to effective excitonic models: a versatile mapping approach including environmental embedding effects}

\abbreviations{DFT, BSE, CT, FE, e-h, TD-DFT, ET, HT, XET, XET}
\keywords{American Chemical Society, \LaTeX}

\begin{document}

\begin{abstract}
We present an original multi-state projective diabatization scheme based on the Green’s function  formalism that  allows the systematic mapping of many-body \emph{ab~initio}
calculations onto effective excitonic models. 
This method inherits the ability of the Bethe-Salpeter equation to describe Frenkel molecular excitons and intermolecular charge-transfer states equally well, as well as the possibility for an effective description of environmental effects in a QM/MM framework. 
The latter is found to be a crucial element in order to obtain accurate model parameters for condensed phases and to ensure their transferability to excitonic models for extended systems. 
The method is presented through a series of examples illustrating its quality, robustness, and internal consistency. 
\end{abstract}


\clearpage
\section{Introduction}
The accurate first-principles description of elementary excitations (excitons) in organic materials represents an invaluable tool for understanding their photophysical properties as well as phenomena underlying applications in optoelectronics, such as energy transfer, exciton splitting or charge separation in organic solar cells.\cite{Scholes2006,Bredas2009,Hetstand2018,Dimitriev2022}

Despite the outstanding progress over the last decades, the \emph{ab~initio} description of excitons in molecular systems still stands as a challenge for theory and computation, because of the difficulty in meeting accuracy and computational cost. A paradigmatic case is represented by the quest for relatively cheap methods able to describe equally well intra-molecular Frenkel-type excitons (FE) and inter-molecular charge transfer (CT) excitation, both characterizing the low-energy region of the excitation spectrum of molecular materials.

Time-dependent density-functional theory (TD-DFT), in its linear-response formulation within the adiabatic approximation,\cite{Casida1995} has been the most popular method for the characterization of excitons in molecular systems. TD-DFT arguably represents one of the most reliable methods among those being computationally affordable for large and complex real-life materials. However, TD-DFT is known to present some limitations, such as the inaccurate and strongly functional-dependent description of CT excitations. 
This results in a possibly incorrect state ordering between CT and FE excitons,\cite{Prlj2015,Stein2009} or in missing the characteristic Coulomb-like dependence of CT states energy with inter-molecular distance when using local, semilocal, or global hybrid functionals.\cite{Dreuw2004} These shortcomings can be significantly attenuated with the use of range-separated functionals,\cite{Yanai2004,Baer2010} leading to results that however depend on the the specific parameterization for the range separation of the Coulomb repulsion term. \cite{Rohrdanz,Chai,Kronik12}

Many-body perturbation methods based on Green's function theory, such as the $GW$\cite{Hedin1965,Str80,Hyb86,Onida2002,ReiningBook,Gol19Rev} and the Bethe-Salpeter equation (BSE) formalisms,\cite{strinati1988application,Bet51,Sha66,Han79} stand as robust and affordable options for computing the excited states in molecular systems.\cite{Rostgaard2010,Kaczmarski2010diabaticBSE,Blase11,Blase11b,Baumeier2012dynamicalBSE,Marom2012,Duchemin12,bruneval2013benchmarking,vanSetten13,Baumeier14,Blase2018}
\revision{Within the resolution-of-the-identity (RI-V) framework,\cite{Duc17_RI} and for converging iteratively a few low-lying BSE eigenstates, our $GW$/BSE implementation  scales as $\mathcal{O}(N^4)$.} 
Besides, $GW$/BSE provides an accurate description of excitations of different natures thanks to a proper treatment of electronic correlations. 
Benchmark studies against reference methods and experimental data demonstrated a typical accuracy of 0.1-0.2~eV for  \revision{charged excitations for valence states and optical transitions, when a self-consistent calculation of quasiparticle energies is performed}.\cite{Faber11,vanSetten2015,Jac15a,Jac15b,bruneval2015systematic,knight2016benchGW,Kaplan2016,Rangel2016,Rangel17,Gui2018benchBSE,Liu2020Thielset,Forster2021} 
\revision{The explicit accounting for the screened
non-local electron-hole (e-h) interaction permits a faithful description of excitations regardless of their nature (FE vs. CT) and of the type of system under scrutiny (e.g. organic homo- and hetero-molecular, inorganic, hybrid).
Importantly, the use of a self-consistent $GW$ scheme largely reduces the dependence of charged and neutral excitations on the functional used as starting point for the many-body perturbation theory.\cite{Blase2018,Jac15a,bruneval2015systematic,Gui2018benchBSE}
We note that a residual starting-point dependence has been  reported for neutral excitations, the magnitude of this fluctuation typically being comparable to  the method's accuracy.\cite{Forster22,Kshirsagar23}
}

The possibility of describing molecular excitations in condensed phases with Green's function many-body method has been enabled by the recent development of a multiscale embedding method of different levels of detail, namely polarizable continuum model (PCM),\cite{Duchemin2016,duchemin2018bethe,clary2023impact,kim2022gw} classical atomistic (polarizable QM/MM)\cite{Li16,Li18,Baumeier14} and full-quantum (QM/QM')\cite{Amblard2023,Amblard2024,sundaram2024quantum}. 
Embedded $GW$ and BSE formalisms have been successfully applied in the contexts of molecular doping\cite{Li17,Comin22,kim2024elucidating,wehner2018electronic} and organic photovoltaics.\cite{Dong2020} Many-body perturbation theory methods have been historically developed for extended (i.e. periodic) systems\cite{Onida2002} and applications to molecular solids have been reported in the literature.\cite{Tiago03,Cudazzo12,Cudazzo13,Cocchi18} 
These calculations are extremely expensive and hardly affordable for realistic molecular solids encompassing many atoms in the unit cell or featuring structural disorder.

\begin{figure}
\centering
\includegraphics[scale=0.5, clip, trim=0.0cm 0.0cm 0.0cm 0.3cm]{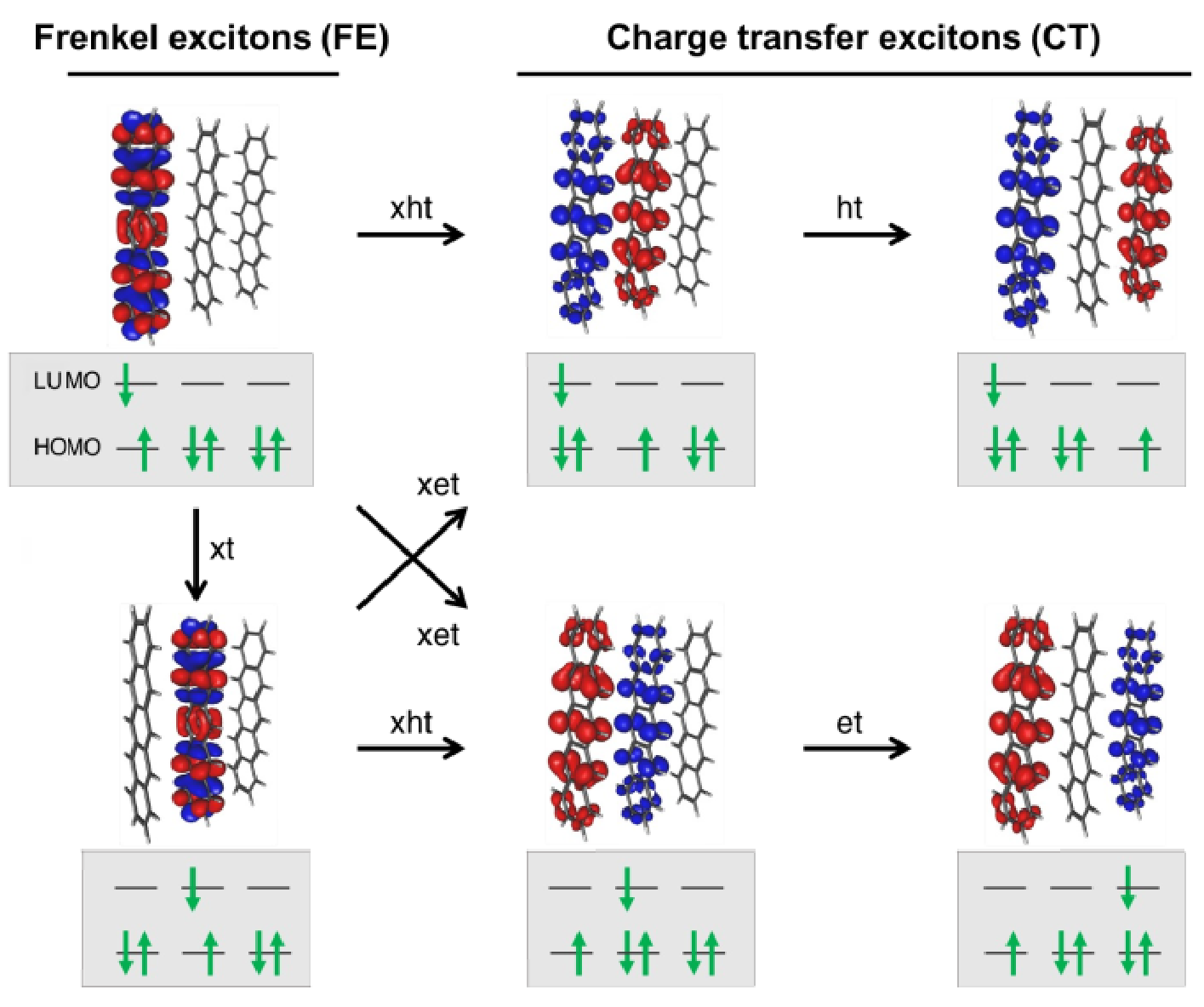}
\caption{Basis diabatic states (pure FE and CT excitons) and main elementary processes defining effective exciton models for supramolecular systems, illustrated for a pentacene trimer extracted from the crystal structure. Hole/electron densities are shown in red/blue isosurfaces of the corresponding probability amplitude. Elementary processes include exciton transfer (XT) and one-body hole/electron transfer coupling FE and CT states (excited-state hole/electron transfer, XHT/XET) and CT states among themselves (hole/electron transfer, HT/ET). Gray boxes display the electronic configurations of basis states. For the sake of illustration, two-level (HOMO, LUMO) molecules have been considered.
}
\label{f:sketch}
\end{figure}

Effective model Hamiltonians for excited states represent a cheap and insightful alternative to a full \textit{ab initio} treatment for the description of the low-energy physics of multichromophoric systems.\cite{MayKuhn,Hetstand2018,Yamagata11,Popp2021}
These approaches rely on the definition of a reduced set of \emph{diabatic} states characterized by wave functions localized on single molecular units, usually including pure intra-molecular FE and pure inter-molecular CT excitations with electron and hole localized on different fragments (see Figure~\ref{f:sketch}). For this reason, these approaches are sometimes referred to as ``low-energy" or ``few-state" models. A typical model Hamiltonian for a molecular solid can be expressed as the sum of three terms,
\begin{equation}
\label{e:H_general}
\textrm{H}=\textrm{H}_{\mathrm{FE}}+  \textrm{H}_{\mathrm{CT}} + \textrm{H}_{\mathrm{CT-FE}}\mathrm{,} 
\end{equation}
describing FEs (standard Frenkel exciton model), inter-molecular CT states, and the interaction between the two types of exciton (see SI for explicit expressions). 
Black arrows in Fig.~\ref{f:sketch} show the elementary processes corresponding to the quantum coupling between states, namely FE  transfer(XT), excited-state hole/electron transfer (XHT/XET), and hole/electron transfer (HT/ET). The energy of the diabatic basis states and the coupling between them represent the parameters entering the FE-CT model Hamiltonian for excited states.
Excitonic models, properly  extended to include the coupling to quantum or classical vibrations find also application in the description of steady-state optical processes\cite{Yamagata11} and real-time dynamics.\cite{Tamura2011,Popp2021,Giannini2022}

The quality of first-principle calculations can be preserved to a large extent by computing the model parameters \textit{ab initio} in the spirit of a multiscale approach.\cite{Yamagata11,Popp2021,Giannini2022} 
For instance, XT couplings between singlet excitons are the only parameters in play in the standard FE model, and several methods have been proposed to its calculations at various levels of theory.\cite{Hsu2009} The simplest approach is the so-called dimer splitting method that, however, strictly applies only to pairs of molecules equivalent by symmetry, each of which presents excitons well-separated in energy. A common approximation consists of neglecting the contributions due to the inter-molecular overlap (exchange terms, e.g. Dexter) and retaining the Coulomb, F\"orster-type, interaction between the transition densities of FEs, which is the leading term at a large inter-molecular distance. Point-dipole and distributed-monopole\cite{Madjet2006} approximations of the transition density offer practical schemes for computing this interaction.

Whenever inter-molecular overlap is of concern, or for more complex FE-CT exciton models, the derivation of parameters from first-principles usually follows so-called diabatization schemes,\cite{Cave96,Rust02,Hsu2008,Subotnik2009,Tamura2011,Difley11,Yang13,Li2015,Arago15,Grofe2017,Green21} in which a unitary transformation is defined transforming the diagonal adiabatic representation of the Hamiltonian of the multichromophoric system into the diabatic one, featuring diabatic-states energies on the diagonal and the couplings among them elsewhere. The choice of the unitary transformation is somewhat arbitrary. Common choices employ the eigenstates of a physically-motivated observable (e.g. in the popular fragment excitation difference scheme, FED\cite{Hsu2008}), or projective schemes where suitable basis functions are introduced.\cite{Difley11,Tamura2011,Green21} 
Projective schemes based on BSE/$GW$ calculations, within the Tamm-Dancoff approximation (TDA),\cite{Tamm1945,Dancoff1950} have been proposed by Wehner and Baumeier\cite{Wehner17,Tirimbo24} and Leng \textit{et al.}.\cite{Leng19} in recent works. 
Also, full and TDA BSE/$GW$ calculations have been recently combined with common diabatization methods (Edmiston-Ruedenberg,\cite{Edmiston1963} generalized Mulliken-Hush,\cite{Cave96} fragment charge difference\cite{Hush1961}) by Baumeier and coworkers.\cite{Tirimbo24}

In this work, we present an original projection-based diabatization scheme based on the BSE/$GW$ formalism, beyond the TDA. The method is designed to be generally applicable to systems with an arbitrary number of excited states of both FE and CT nature. A distinctive feature of our methodology is the employ of QM/MM embedding techniques for building effective excitonic models for systems in condensed phases. Special emphasis is put on the transferability of the parameters obtained for small molecular clusters (dimers or trimers) to excitonic models for extended systems. The use of classical MM embedding turns out to be essential for such a purpose. Overall, the proposed methodology provides a general, robust, and versatile framework for the systematic mapping of many-body \textit{ab~initio} calculations to effective excitonic models for large supramolecular systems such as solids, films, and interfaces. 

\section{Theory}
\label{s:theory}

\subsection{$GW$ and Bethe--Salpeter equation formalism}
\label{s:gwbse}

The present development aims to set up 
effective excitonic models within the framework of the many-body \emph{ab initio} Green's function techniques $GW$ and BSE. 
This subject is briefly outlined here from a practical perspective, highlighting the features that will be mostly relevant to this end. 
For further detail, we refer the readers to review papers\cite{Onida2002,Blase2018,ReiningBook,Gol19Rev} and to the original publications of our implementation.\cite{Blase11,Blase11b,Duchemin2016,Li16,Li18} 

Methods based on Green's functions rely on many-body perturbation theory to compute electronic excitations upon introducing electronic correlation effects on top of mean-field electronic structure calculations, such as Kohn-Sham DFT. \cite{KohnSham65}
The set of Kohn-Sham orbitals $\{\phi^\mathrm{KS}\}$ and one-particle energies $\{\varepsilon^\mathrm{KS}\}$ obtained with a given density functional approximation define the non-interacting system  used as a starting point for the perturbation theory. 
Then, within this framework, charged excitations such as electron removal (e.g. ionization potential) and electron addition (e.g. electron affinity) can be computed as quasiparticle energies
\begin{equation}
 \varepsilon_n = \varepsilon^\mathrm{KS}_n + 
\langle \phi_n |  \Sigma(\varepsilon_n) - V_{xc} | \phi_n \rangle,
\label{eq:gw}
\end{equation}
where, in practice, the self energy $\Sigma(\varepsilon_n)$ replaces the Kohn-Sham DFT exchange-correlation potential $V_{xc}$. 
The self-energy is a non-local and energy-dependent operator, $\Sigma({\bf r},{\bf r}';\varepsilon)$.
This, within the $GW$ formalism, is approximated as 
the product of the single-particle Green's function ($G$) and the screened Coulomb interaction ($W$),
i.e. $\Sigma^{GW}=\textrm{i}GW$ with i being the imaginary unit.
In the so-called $G_0W_0$ scheme, $\Sigma^{GW}$ is built directly with the Kohn-Sham levels. Re-computing the self-energy with quasiparticle energies iteratively until convergence leads to the partially self-consistent ev$GW$ scheme.
This approach largely reduces the dependence of quasiparticle excitations on the starting DFT functional, leading to a  good agreement with experimental values or high-level CCSD(T) calculations.\cite{Faber11,Rangel17,Kaplan16}

Neutral excitations can be obtained \emph{via} the BSE, which is a two-body correlated e-h problem described by the following non-Hermitian eigenvalue equation
\begin{equation}
\left(\begin{array}{cc}
\mathbf{A} & \mathbf{B} \\
\mathbf{B}^* & \mathbf{A}^*
\end{array}\right) 
\left(\begin{array}{c}
\mathbf{X}_k \\
\mathbf{Y}_k
\end{array}\right)=\Omega_k\left(\begin{array}{c}
\mathbf{X}_k \\
\mathbf{Y}_k
\end{array}\right)
\label{e:bse}
\end{equation}
where $\Omega_k$ is the excitation energy of the $k$-th state, the corresponding eigenvector including an excitation ($\mathbf{X}_k$, occupied~$\rightarrow$~virtual) and a de-excitation ($\mathbf{Y}_k$, virtual~$\rightarrow$~occupied) component.
For singlet excitations, the resonant and anti-resonant blocks of the BSE Hamiltonian reads:
\begin{eqnarray}
\label{e:bse_AB}
A_{ai,bj} &=&
 \delta_{ab} \delta_{ij} 
 \left(\varepsilon_a^{G W}-\varepsilon_i^{G W}\right)+2\left\langle\phi_a(\rv) \phi_j\left(\rv^{\prime}\right) | v\left(\rv, \rv^{\prime}\right) | \phi_i(\rv)  \phi_b\left(\rv^{\prime}\right) \right\rangle 
  \nonumber \\
&-&\left\langle\phi_a(\rv ) \phi_j\left(\rv^{\prime}\right) | W\left(\rv, \rv^{\prime},\omega=0\right) | \phi_b(\rv) \phi_i\left(\rv^{\prime}\right) \right\rangle \\
B_{ai,bj} &=&2\left\langle\phi_a(\rv) \phi_b\left(\rv^{\prime}\right) | v\left(\rv, \rv^{\prime}\right) | \phi_i(\rv)     \phi_j\left(\rv^{\prime}\right) \right\rangle 
   \nonumber \\
&-&\left\langle\phi_a(\rv )\phi_b( \rv^{\prime} )  | W\left(\rv, \rv^{\prime},\omega=0\right) |  \phi_j\left(\rv \right)   \phi_i\left(\rv^{\prime}\right)  \right\rangle
\end{eqnarray}
%
where $v({\bf r},{\bf r}')=1/|\rv -{\rv}'|$ is the bare Coulomb interaction and $W$ is the screened one in the static limit ($\omega=0$). 
Let us comment that the Kohn-Sham orbitals $\lbrace \phi_n \rbrace$ are real-valued in the present work. The anti-resonant blocks $\mathbf{B}$ and $\mathbf{B}^*$ 
are neglected in the Tamm-Dancoff approximation~\cite{Tamm1945,Dancoff1950}, which is not adopted in this work.
We emphasize that the screened-Coulomb potential matrix elements in the resonant block $\mathbf{A}$ do not cancel  when occupied $(i,j)$ states do not overlap with unoccupied $(a,b)$ states. 
This explains the success of BSE for CT excitations. 

BSE eigenvectors are normalized (i.e. $\sum_{ia} \left[ (X_k ^{ia})^2-(Y_k ^{ia})^2 \right] =1$) and orthogonal to each other, 
allowing us to define the two-body (electron-hole)  exciton wavefunction as
\begin{equation}
\Psi_k (\rvh,\rve ) = 
\sum_{ia}\left\{X_k^{ia} \phi_i ( \rvh ) \phi_a (\rve)+
Y_k^{ia} \phi_i\ (\rve ) \phi_a (\rvh )\right\}
\label{e:bse_wf}
\end{equation}
and the corresponding transition density
\begin{equation}
n_k ^{\textrm{T}} ({\bf r}) = \sqrt{2}
\sum_{ia}(X_k^{ia} + Y_k^{ia}) \phi_i ( {\bf r}) \phi_a ({\bf r}).
\label{e:bse_tdens}
\end{equation}

We recall that, despite the similarity between equation~\ref{e:bse} and the popular Casida's linear-response equation in TD-DFT,\cite{Casida1995} the explicit account for the \revision{screened} e-h interaction permits a faithful description of excitations, regardless of their nature (FE or CT) or of the Kohn-Sham-DFT density functional approximation employed to construct the single-particle basis. This determines an important qualitative and quantitative improvement over TD-DFT, especially for the stability of CT excitation energies, making BSE a robust and predictive tool for describing FE and CT excitons on an equal footing.\cite{jacquemin2017benchmark}

\subsection{Environmental embedding  in $GW$/BSE}
\label{s:environ}

Many-body calculations of molecules and small molecular clusters in condensed phases have been performed by adopting hybrid quantum mechanics/classical mechanics (QM/MM) embedding techniques.\cite{Duchemin2016,Li16,Li18,Blase2018} 
These approaches require accounting for the effect of the embedding environment at different steps of the computational workflow, first at the ground-state Kohn-Sham DFT level and then in the calculation of charged and neutral excitations.

Environmental embedding in the ground state is essential for molecules dissolved in polar solvents, as well as in the solid state. 
Ground-state electrostatic embedding is indeed necessary to describe the strong crystal fields sourced by molecular multipole moments.
A proper description of internal fields is essential for the quantitative description of the absolute binding energies in photoemission experiments on molecular solids,\cite{Li18} as well as to capture the dependence of energy levels on molecular orientations at film and crystal surfaces.\cite{Dong2020} 
In the following, ground-state embedding will be either based on PCM,\cite{Duchemin2016} to describe a generic non-polar medium, or on the charge response polarizable atomistic model,\cite{Tsi01,Dav14} whenever molecular crystals are the explicit target.

The treatment of the screening of charged ($GW$) and neutral (BSE) excitations in the presence of an environment has been documented in our original works.\cite{Duchemin2016,Li16,Li18} 
The key for this development consists in treating the dielectric susceptibility of the embedding medium as a frequency-independent quantity, i.e. the environment reacting instantaneously to the excitations within the QM region. 
The frequency-independent response of the environment is fully consistent with the BSE, which also relies on a static screening of the e-h interaction (see Eq.~\ref{e:bse_AB}). 
The  susceptibility is instead an explicitly dynamic quantity in the $GW$ formalism, requiring an additional step for combining the dielectric response of QM and MM regions. 
The effect of the screening of charged excitations (polarization energy) can be computed at the static COHSEX level, i.e. the frequency-independent analog of $GW$.\cite{Hedin1965} 

In previous works, the state-specific polarization energy was computed as the difference between the quasiparticle energy obtained in two independent COHSEX calculations including or not the effect of the environment. Then, this difference was added as a perturbative correction to $GW$ quasiparticle energies ($\Delta$COHSEX).\cite{Li16,Li18} 
An alternative treatment of the environment emerged in the context of a QM/QM' framework based on a fragment formalism.\cite{Amblard2024}
This approach allows an explicit account for the dynamics of the embedding medium (frequency $\omega_\mathrm{env}$), but also to recover the instantaneous dielectric response of the environment in the $\omega_\mathrm{env}\rightarrow\infty$ limit. This permits the description of the dielectric embedding beyond the perturbative $\Delta$COHSEX approach. 
The novel self-consistent COHSEX treatment of environmental screening has been combined with PCM and adopted in the present work. The details of the methodology will be disclosed in a forthcoming publication.

A central quantity in embedded $GW$/BSE is the so-called reaction field matrix, 
$v_\mathrm{reac}(\rv,\rv^{\prime})$, 
describing the electric potential generated by the environment at a probe point $\rv$ in response to a source charge at $\rv^\prime$.
The reaction field matrix can be written as
\begin{equation} \label{e:v_reac} 
v_\mathrm{reac}(\rv,\rv^\prime) = 
\int d\rv_2 d\rv_2^\prime  \; 
v(\rv,\rv_2) \chi_\mathrm{MM} (\rv_2,\rv_2^\prime)   v(\rv_2^\prime,\rv^\prime),
\end{equation}
where $\chi_\mathrm{MM} (\rv_2,\rv_2^\prime)$ is the interacting susceptibility of the environment alone.
For PCM embedding, Eq.~\ref{e:v_reac} corresponds to the interaction of a probe charge with the charge density induced by the source charge on the surface of the cavity. In embedded $GW$ and BSE calculations, the MM environment renormalizes the Coulomb potential within the QM region, i.e. $v \leftarrow v + v_\mathrm{reac}$. 

In summary, in the following, we will perform embedded DFT/$GW$/BSE calculations using different schemes. 
PCM embedding has been used at every level of the calculation for model dimers reported in Section~\ref{s:dimers}. 
A hybrid scheme has been instead adopted for structures extracted from crystal structures, see Section~\ref{s:crystals}. In this case, an atomistic embedding has been adopted in the DFT calculation, to ensure a proper description of crystal fields in the neutral ground state, while the screening of excitations has been treated at the PCM level. The latter should be considered as an approximation of the anisotropic dielectric response of the crystal that is adopted for practical reasons. 

\subsection{Projective diabatization method}
\label{s:scheme_new}

Given a reduced set of physically-motivated excitonic basis functions including FE and CT states, our aim is to obtain the matrix representation of an effective model Hamiltonian such that it reproduces excitation energies and wavefunctions of a BSE calculation on a small supramolecular cluster.
Diabatic basis functions with excitons and charges localized on single  molecular units (fragments) can be normalized but are not orthogonal to each other, requiring accounting for their overlap in a generalized eigenvalue problem.
Keeping into explicit account the overlap among basis functions is actually an important ingredient for our long-term goal, which is to build  models for large supramolecular clusters or extended systems in a modular fashion, namely using matrix elements derived only from BSE calculations on dimers or small aggregates. 
This procedure further allows one to systematically expand the system by simply adding additional basis functions to our set and deriving the corresponding effective Hamiltonian and overlap matrix elements without having to reconsider the coefficients already at hand.

We consider a system composed of distinct molecular fragments, where we perform BSE calculations on the single fragments and for the whole system. 
For the sake of simplicity, a system composed of two fragments (dimer) is considered in the following. The generalization to ensembles of more fragments is straightforward.
We define a set of e-h states $\{\psi\}$ that will be used as a diabatic basis to set up the exciton-model Hamiltonian, including FE excitons localized  on one of the two fragments and inter-fragment CT states.
Diabatic FE states are set to the BSE eigenstates of individual fragments, with $\psi_l^\mu=\Psi_l^\mu(\rvh,\rve)$ corresponding to the $l$th exciton of fragment $\mu$.
CT basis states are defined as 
$\psi_{ia}^{\mu,\nu} (\rvh,\rve) = {\phi_{i} ^\mu}(\rvh) {\phi_{a} ^\nu}(\rve)$ with $i$ referring to an occupied Kohn-Sham orbital of fragment $\mu$ and $a$ to a virtual state of fragment $\nu$. 
The number of selected FE and CT basis states (problem dimension $D$, fixed by the range spanned by $l$, $a$ and $i$) can be determined \emph{a posteriori}, based on the energy range the model is expected to cover \revision{(see below)}.

To build the low-energy Hamiltonian, we first introduce the overlap matrix between basis functions $\psi_p$ (hereafter identified with a single index $p=1,2\dots D$)
\begin{equation} \label{e:S}
 S_{pq}= \langle \psi_p | \psi_q \rangle, 
\end{equation}
and the projection matrix
\begin{equation} \label{e:P}
 P_{kq}= \langle  \Psi_k | \psi_q \rangle
\end{equation}
where $\Psi_k$ are the BSE eigenstates of the dimer with associated eigenvalues $\Omega_k$. 
The target eigenstates for the model Hamiltonian $\widetilde \Psi_k$, corresponding to the projection of BSE wavefunctions in the vector space spanned by the basis, are  expressed as a linear combination of the diabatic states,
\begin{equation}
\widetilde \Psi_k =\sum_p C_{pk}\ \psi_p,
\end{equation}
the matrix of the expansion coefficients being
\begin{equation} \label{e:coeff}
\mathbf{C}={\bf S}^{-1} \mathbf{P} \boldsymbol{\Lambda}.
\end{equation}
This expression directly follows from the application to $\widetilde \Psi_k$ of the identity operator, in the form of a completeness relation for a non-orthogonal basis.
The diagonal matrix $\boldsymbol{\Lambda}$, with elements 
$\Lambda_{kj}=\left[{({\bf P}^\textrm{T}{\bf S}^{-1}{\bf P})_{kk}}\right] ^{-1/2}\delta_{kj}$, 
ensures the normalization of the target functions, but not their orthogonality.  
We note that while the BSE wavefunctions of the dimer ($\Psi_k$) form an orthonormal set, the target eigenstates for the model Hamiltonian ($\widetilde \Psi_k$) do not.
Starting from such a common background, two methods can be proposed to obtain the Hamiltonian. 

A straightforward approach consists of obtaining the Hamiltonian directly from the generalized eigenvalue equation as 
\begin{equation} \label{e:H1}
\mathbf{H}=\mathbf{SC}\boldsymbol{\Omega}\mathbf{C}^{-1}
\end{equation}
where $\boldsymbol{\Omega}$ is the diagonal matrix of the BSE eigenvalues. 
However, because of the non-orthogonality of $\tilde \Psi_k$ functions, the resulting $\mathbf{H}$ matrix is not symmetric as it should be.
When  asymmetry is small, as expected from a reasonably large basis set, a practical workaround could be the symmetrization by hand, i.e. $\mathbf{H} \leftarrow (\mathbf{H} + \mathbf{H}^\mathrm{T})/2$, where ${\bf H}^\textrm{T}$ is the transpose of ${\bf H}$. 

Alternatively, we may recognize that the transpose of the Hamiltonian should also lead to the same eigenvalue problem (i.e. ${\bf H}^\textrm{T} {\bf C}={\bf SC}{\boldsymbol{\Omega}} $). 
Therefore, we can formulate this as the optimization problem of finding a symmetric Hamiltonian ${\bf H}$ as
\begin{equation} \label{e:Hmin}
    \min_{{\bf H}} \left[ || {\bf H C}-{\bf SC}{\boldsymbol{\Omega}}  ||+|| {\bf H}^\textrm{T} {\bf C}-{\bf SC}{\boldsymbol{\Omega}} || \right],
\end{equation}
where $||{\bf Z}|| = \textrm{Tr}({\bf Z}^\textrm{T} {\bf Z})$ and $\textrm{Tr}$ denotes the trace of the matrix ${\bf Z}$. 
Then, expressing Eq.~\ref{e:Hmin} in terms of traces of product of matrices and following the usual rules for the derivatives of matrices~\cite{klein1999calculus}, we arrive at
\begin{equation}
\label{e:sylv}
{\bf CC}^\textrm{T} {\bf H}  +  {\bf H} {\bf CC}^\textrm{T} = {\bf C} {\boldsymbol{\Omega}} {\bf C}^\textrm{T} {\bf S} + {\bf S C }{\boldsymbol{\Omega}} {\bf C}^\textrm{T}
\end{equation}
that corresponds to a Sylvester's equation~\cite{hu1992krylov},
which allows us to find the Hamiltonian ${\bf H}$ that is symmetric and best reproduces BSE eigenvalues and the target functions.
Equation~\ref{e:sylv}  can be solved with the Bartels–Stewart algorithm,\cite{bartels1972algorithm} see SI for  details.

\revision{Before closing this section, we point out some practical considerations regarding the selection of the number and nature of basis states. 
The construction of a low-energy model relies on the somewhat arbitrary choice on the energy cutoff between the degrees of freedom to include in the model and those to discard.
This choice must be physically motivated, it should reduce as much as possible the dimension of the problem, but also   
take into account common sense arguments. 
For instance, the energy gap between the highest-energy target BSE exciton and the first-excluded one should be sizeable.
We advise against cutting through manifolds of degenerate or quasi-degenerate states.\\
The number of basis states $D$ must then equal the number of target states.
The $P$ matrix in Eq.~10 can be first evaluated for a larger set of candidate  states, possibly including multiple FE and CT excitons per fragment. 
The inspection of $P$ permits to identify the $D$ states that would finally constitute the basis, i.e. the state for which the projection of the dimer BSE excitons is the largest. 
For example, the minimal basis for molecular homodimers consists in one FE per fragment, albeit in some cases the inclusion of low-lying CT states might be appropriate, as in the pentacene case study developed below.
In molecular heterodimers, such as donor-acceptor complexes, the inclusion in the basis of CT transitions from the occupied levels of the donor to the unoccupied subspace of the acceptor is crucial. }

\subsection{\revision{Approximate excitonic couplings}}
\label{s:approx_v}

Besides the  projective diabatization scheme described above, we also compute exciton couplings within common approximations. 
The FE-CT and some of the CT-CT couplings can be approximated as the matrix elements of the Kohn-Sham Hamiltonian between frontier molecular orbitals of the involved fragments.
Referring to the example in Figure~\ref{f:sketch} for two-level molecules (HOMO, LUMO), the one-electron  approximation for the coupling associated with XHT and HT  process between fragments $\mu$ and $\nu$ is 
$\langle \phi_\mathrm{HOMO}^\mu |\widehat{\textrm{H}}_{\textrm{KS}}^{\mu\nu}|\phi_\mathrm{HOMO}^\nu\rangle$,
where $\widehat{\textrm{H}}_{\textrm{KS}}^{\mu\nu}$ is the Kohn-Sham Hamiltonian for the dimer $\mu$-$\nu$ and 
$\phi_\mathrm{HOMO}^\mu$ is HOMO of fragment $\mu$.
Similarly, the matrix element for XET and ET processes reads $\langle \phi_\mathrm{LUMO}^\mu |\widehat{\textrm{H}}_{\textrm{KS}}^{(\mu\nu)}|\phi_\mathrm{LUMO}^\nu\rangle$.

In the limit of zero intermolecular overlap, the coupling between singlet FE excitons reduces to the electrostatic interaction between transition densities.\cite{MayKuhn}
For a pair of molecules in the vacuum the FE-FE coupling is
\begin{equation}\label{e:dens_dens_V}
V^{n-n}_\mathrm{FE-FE} (n_\mu ^{\textrm{T}}, n_\nu ^{\textrm{T}} )
= \int d{\bf r}d{\bf r}' n_\mu ^{\textrm{T}} ({\bf r}) 
v({\bf r},{\bf r}') n_\nu ^{\textrm{T}} ({\bf r}')
\end{equation}
where $n_\mu ^{\textrm{T}}$ is the transition density of a given excitation on fragment $\mu$.
Following a multipole expansion of $n_\mu ^{\textrm{T}}$ truncated to the leading term, one obtains the coupling as the interaction between transition dipoles (${\bf d}_{\mu} ^{\textrm{T}}$) as
\begin{equation}\label{e:dip_dip_V}
V^{\mathrm{d-d}} _\mathrm{FE-FE} = \frac{|{\bf d}_\mu ^{\textrm{T}}|\;|{\bf d}_\nu ^{\textrm{T}}|}{|{\bf R}_{\mu\nu}|^3} \left[\widehat{{\bf d}}_\mu ^{\textrm{T}} \cdot \widehat{{\bf d}}_\nu ^{\textrm{T}}-3( \widehat{{\bf d}}_\mu ^{\textrm{T}} \cdot \widehat{\bf R}_{\mu\nu} )( \widehat{{\bf d}}_\nu ^{\textrm{T}} \cdot \widehat{\bf R}_{\mu\nu} ) \right],
\end{equation}
where ${\bf R}_{\mu\nu}$ is the inter-fragment distance and hat symbol denotes unit vectors.
The dipolar approximation holds at a large distance, with a characteristic $|{\bf R}_{\mu\nu}|^{-3}$ decay of the coupling.

\subsection{Dielectric screening of FE-FE couplings}
\label{s:screen_Vxt}

\revision{The presence of a dielectric environment (e.g. PCM) largely affects  FE-FE couplings.
Restricting ourselves to the leading electrostatic interaction between transition densities (Eq.~\ref{e:dens_dens_V}), the embedding medium affects the coupling in two ways, as discussed by Mennucci and coworkers. \cite{scholes2007solvent,curutchet2007solvent} }
The first trivial effect, results from the change in the transition density of each molecular fragment due to the embedding polarizable medium.
The second contribution is due to the dielectric screening of the field exerted by the transition densities.
Both effects are considered in our calculations, namely, transition densities are obtained from embedded BSE calculations and  screened couplings are computed as
\begin{equation}\label{e:dens_dens_V_env}
{V}^{n-n,\mathrm{e}} _\mathrm{FE-FE} 
(n_\mu ^{\mathrm{T,e}}, n_\nu ^{\mathrm{T,e}} )
= \int d{\bf r}d{\bf r}'\  n_\mu ^{\mathrm{T,e}} ({\bf r}) 
\left[ v({\bf r},{\bf r}') + v_{reac}({\bf r},{\bf r}') \right]
n_\nu ^{\mathrm{T,e}} ({\bf r}'),
\end{equation}
where the superscript ``e" labels quantities calculated in the environment and 
$v_{reac}({\bf r},{\bf r}')$ is the medium reaction field defined in Equation~\ref{e:v_reac} that here is calculated at the PCM level.
Following Mennucci and coworkers,\cite{curutchet2007solvent,scholes2007solvent} we introduce an effective screening factor
\begin{equation} \label{e:s_eff}
s=\frac{ V^{n-n,\mathrm{e}}_\mathrm{FE-FE}(n_\mu ^{\mathrm{T,e}}, n_\nu ^{\mathrm{T,e}} ) }{V^{n-n}_\mathrm{FE-FE}(n_\mu ^{\mathrm{T,e}}, n_\nu ^{\mathrm{T,e}} )}   
\end{equation}
\revision{where the screened and unscreened interactions are both calculated for the same transition densities, here obtained in PCM.
Computing the ratio for the same transition density is important to properly disentangle the two effects of the polarizable medium on the excitonic coupling, namely the impact of the environment on the transition density and the dielectric screening of the interaction.}

The screening factor $s$ can be evaluated for any transition density and polarizable embedding model.
It is interesting to consider the case of point transition dipoles in a continuum dielectric.
For dipoles incorporated in the dielectric, without an explicit cavity, the screening factor is $s=1/\epsilon_\mathrm{opt}$.
In the limit of the large distance between the two dipoles enclosed into spherical cavities, the screening factor reads
\begin{equation}\label{e:s_eopt} 
s=\epsilon_\mathrm{opt} \left(\frac{3}{1+2 \epsilon_\mathrm{opt}}\right)^2
\end{equation}
where $\epsilon_\mathrm{opt}$ is the relative permittivity in the optical range.
The derivation of this expression is provided as Supporting Information.
To the best of our knowledge, this result has not been reported before. 
We have explicitly verified the agreement between  Equation~\ref{e:s_eopt} and numerical calculations based on our PCM implementation.\cite{Duchemin2016} 
For dipoles located at the cavity center, the limiting values are quickly recovered for dipole-dipole distances larger than twice the cavity radius, i.e. once the two dipoles are embedded in separate cavities.
We emphasize that two elements concur to the result in  Equation~\ref{e:s_eopt}: 
(i) the screening of the field of one of the dipoles (source dipole) by the charges induced on the surface of its cavity;
(ii) the Onsager cavity field generated within the cavity hosting the probe dipole by the (screened) field of the source dipole.\cite{onsager1936electric}

\subsection{Computational details}
\label{s:details}
In this work, we performed PBE0 calculations~\cite{Perdew96,adamo1999toward} using the \texttt{ORCA} package~\cite{neese2022software} to obtain the Kohn-Sham wavefunctions. 
$GW$ and BSE calculations were carried out with the \texttt{BeDEFT} program.\cite{duchemin2021cubic}
All calculations were done with the def2-TZVP basis set~\cite{weigend2005balanced} a the corresponding def2-TZVP-RI auxiliary basis for Coulomb fitting\cite{weigend1998ri} in the resolution-of-the-identity framework (RI-V).\cite{Duc17_RI}
ev$GW$ calculations have been performed, correcting 4, 8, and 12 states in calculations of single fragments, dimers, and trimers, respectively. 
For PCM embedding, we adopted a dielectric constant $\epsilon_r=\epsilon_\mathrm{opt}=3.5$, typical of non-polar organic solids. 
Other parameters were set to the default values for benzene from the Minnesota Solvent Descriptor Database.
DFT calculations employed the ORCA-code implementation of the conductor-like polarizable continuum model (C-PCM).\cite{Barone98}
A double-layer formulation of the 
integral equation formalism PCM model (IEF-PCM), allowing for charge-spilling effects, has been used in $GW$ and BSE computations.\cite{Duchemin2016}
The ground-state embedding of calculations on the pentacene crystal employed the charge response model\cite{Tsi01} as implemented in the MESCal code.\cite{Dav14}
The pentacene molecular geometry employed in model dimer calculations (Section~\ref{s:dimers}) has been optimized at the B3LYP/def2-TZVP level in the gas phase.
Dimers have been then built by creating a translated replica of the molecule along the direction perpendicular to the molecular plane, obtaining parallel co-facial dimers for different intermolecular distances.
Calculations on the pentacene crystal (Section~\ref{s:crystals}) have been performed for  the Siegriest triclinic structure\cite{Sie01}.

\section{Results}
\label{s:results}

We now present some illustrative examples of the application of our methodology for deriving effective excitonic models from many-body BSE/$GW$ calculations.
Pentacene is taken as an ideal test case given its prominent role in organic electronic research. 
The following test cases allow us to discuss the merits of our approach, from the perspective of applying our methodology to supramolecular systems in the condensed phase.
Selected examples include molecular homodimers in the vacuum and PCM, as well as the pentacene molecular crystal.

Effective model Hamiltonians are obtained with the diabatization method based on the optimization approach (solution of Equation~\ref{e:sylv}). 
We emphasize that the diagonalization of the model Hamiltonian allows reproducing with high precision the target eigenvalues ($\Omega_k$) and eigenvectors ($\widetilde\Psi_k$) from BSE calculations.
We have also considered the Hamiltonian obtained \emph{via} Equation~\ref{e:H1} and subsequent symmetrization by hand.
The differences in the eigenvalues of the model $\mathbf{H}$ obtained with the two methods are typically below 1 meV, which is practically negligible if compared to the approximation inherent to a few-state model or the intrinsic precision of BSE/$GW$ calculations. 
The optimization method is preferred, being a more rigorous and elegant approach.

\subsection{Model  molecular dimers}
\label{s:dimers}

\subsubsection{Pentacene homodimer in vacuum}
\label{s:pen}

We first analyze the case of a molecular homodimer in the vacuum.
We consider molecular pairs arranged in a parallel cofacial geometry, with the intermolecular stacking distance $R$ scanned from 3.5 to 15 \AA.
The primary excitations of pentacene homodimers can be described in terms of two FE excitations, i.e. the $S_1$ excitons of the isolated fragments, and two CT states, corresponding to HOMO$\to$LUMO intermolecular transitions.
Our basis to set up the model $\mathbf{H}$ is set accordingly, resulting in a 4-state model.

A distinctive aspect of our methodology is that a single (DFT, ev$GW$, BSE) calculation is performed for one of the two symmetry-equivalent molecules in the dimer, the second being created as a copy of the first one. 
This means that the atomic orbitals, Kohn-Sham orbitals, and BSE wavefunctions of  fragment 2 are a translated replica of those of fragment 1. 
This fixes a  convention for the phase of the basis excitonic functions that define the sign of the off-diagonal terms of $\mathbf{H}$ (excitonic couplings).
This will be particularly important for calculations on crystals, as discussed later in Section~\ref{s:crystals}.

\begin{figure}[H]
 \includegraphics[scale=1]{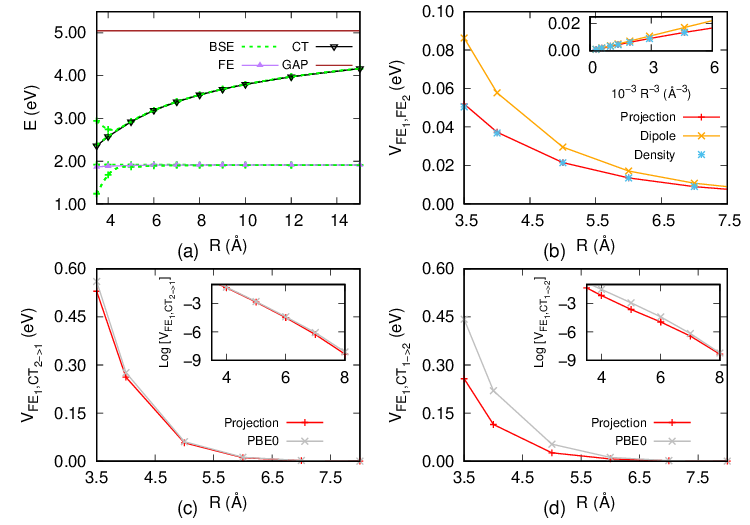}
 \caption{Results of the projective diabatization method for a pentacene dimer in the gas phase as a function of the inter-fragment distance $R$.
 (a) Diabatic energies  compared to the BSE transition energies and to the ev$GW$ HOMO-LUMO gap.
 (b) FE-FE  couplings as obtained with the diabatization scheme and as Coulomb interaction between transition densities and transition dipoles.
 (c-d) FE-CT couplings (absolute values) including the approximate one-electron matrix elements obtained at the PBE0 level.
 Panels a,b,c report the absolute value of the couplings between excited states. }
 \label{fig:pentacene_gas}
\end{figure}

Having defined the basis, BSE calculations can be performed on the dimer to obtain the excitonic model Hamiltonian according to the procedure described in Section~\ref{s:scheme_new}.
For example, the upper-triangular part of the Hamiltonian (in eV) at $R=5$ \AA\  one obtains
\begin{equation}
\label{e:Hpp}
{\bf H}_\mathrm{pen-pen} =\ \ 
\begin{NiceArray}{rrrrl}
\mathrm{FE}_1 & \mathrm{FE}_2 & \mathrm{CT}_{1\to2} & \mathrm{CT}_{2\to1} \\
1.898 & 0.022 & 0.026 &  -0.057 & \mathrm{FE}_1 \\
      & 1.898 &  -0.057 & 0.026 & \mathrm{FE}_2 \\
      &       &  2.923 &  0.000 & \mathrm{CT}_{1\to2}\\
      &       &        &  2.923 & \mathrm{CT}_{2\to1} \\
\CodeAfter 
  \SubMatrix({2-1}{5-4})      
\end{NiceArray}
\end{equation}
which upon diagonalization yields eigenvalues that reproduce the 4 lowest-energy singlet BSE excitation energies (i.e. 1.870, 1.920, 2.924, and 2.930 eV) within 1 meV tolerance. 
The diagonal elements in Eq.~\ref{e:Hpp} represent the diabatic energies that are plotted for different inter-fragment distances in Fig.~\ref{fig:pentacene_gas}a, together with the BSE energies and the ev$GW$ HOMO-LUMO gap. 
We observe that the diabatic energies follow the behavior expected from fundamental physical considerations.
The diabatic energies of the two localized FEs remain constant with the inter-fragment distances, rapidly approaching the two degenerate FEs of the dimer in the $R\to\infty$ limit.  
The diabatic energy of CT states follows the $1/R$ behavior prescribed by the e-h interaction for these intermolecular excitons, converging toward the photoemission gap at large $R$.
The diabatic energies converge to the adiabatic dimer excitation energies for $R\to\infty$, while excitonic couplings determine significant deviations at distances typical of intermolecular contacts.

The off-diagonal elements of the Hamiltonian correspond to FE-FE, FE-CT, and CT-CT couplings. 
The FE-FE couplings obtained at different inter-fragment distances are displayed in Fig.~\ref{fig:pentacene_gas}b. 
This plot shows that the coupling decays with the inter-fragment distance as $R ^{-3}$ (see inset), as expected in the limit of interacting transition dipoles.
We further observe that the coupling obtained from the projective diabatization matches well with the Coulomb interaction between transition densities (zero-overlap F\"orster limit, Eq.~\ref{e:dens_dens_V}) down to short contact distance.
The dipolar approximation (Eq.\ref{e:dip_dip_V}) is  recovered  in the $R\to\infty$ limit.
FE-CT couplings are shown in Fig.~\ref{fig:pentacene_gas}c,d, featuring an exponential decay in $R$ (see insets).
Small deviations from a simple exponential dependence at $R>6$~\AA\ might be ascribed to the use of a Gaussian basis functions, instead of Slater-type ones. 
For FEs that are pure intra-fragment HOMO-LUMO transitions, these couplings can be approximated with one-electron matrix elements between Kohn-Sham orbitals (see Section~\ref{s:approx_v}), namely 
$\langle \phi^1 _{\textrm{HOMO}}|\widehat{\textrm{\textrm{H}}}_\textrm{KS}|\phi^2_{\textrm{HOMO}}\rangle$ for FE$_1$-CT$_{2\to1}$ coupling (panel c) and
$\langle \phi^1 _{\textrm{LUMO}}|\widehat{\textrm{\textrm{H}}}_\textrm{KS}|\phi^2_{\textrm{LUMO}}\rangle$ for  FE$_1$-CT$_{1\to2}$ coupling (panel d). 
The coupling computed with PBE0 functional captures the correct distance decay, resulting from intermolecular overlap, but might differ quantitatively from the diabatization result based on BSE.
Specifically, we find that the PBE0 approximate couplings lie close to the ones obtained by the projection method for the HOMO-HOMO interaction while they present some deviations for the LUMO-LUMO ones.
This suggests that many-body effects might have a significant weight in the evaluation of these matrix elements.
Finally, we observe that the direct coupling between CT states (see Eq.~\ref{e:Hpp}) is negligible for $R>$5 \AA, because the corresponding process, CT$_{1\to2}\to$CT$_{2\to1}$, is a two-electron hop connecting barely overlapping  diabatic states.

\subsubsection{Pentacene homodimer in PCM}
\label{s:pentacene_pcm}

We next consider the effect of a polarizable environment, by computing the same pentacene dimer, as a function of $R$, in PCM.
We opt for a dielectric medium with typical permittivity of a nonpolar organic material $\epsilon_r=\epsilon_\mathrm{opt}=3.5$. 
The calculation setup follows the same procedure previously described for the gas-phase case, except that the calculations on the single fragment, needed to build the basis, and those on dimers are both performed in PCM.
Two disconnected PCM cavities occur in dimer calculations at large intermolecular distances, merging into a single one when molecules come into contact.

\begin{figure}[H]
\includegraphics[scale=1]{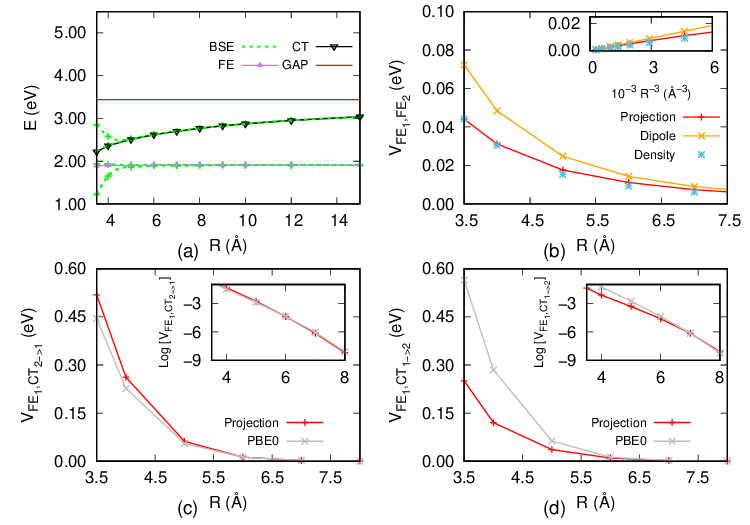}
 \caption{Results of the projective diabatization method for a pentacene dimer in PCM ($\epsilon_r=3.5$) as a function of the inter-fragment distance $R$.
 Results are presented in Figure~\ref{fig:pentacene_gas}.
 The main differences introduced by PCM embedding are the 1.6~eV reduction of the gap due to screening phenomena and the stabilization of diabatic CT states in panel a.
 The magnitude of FE-FE couplings (panel b) is also reduced in magnitude by dielectric screening. }
 \label{fig:pentacene_pcm}
\end{figure}

The results of our diabatization scheme are reported in Figure~\ref{fig:pentacene_pcm} as a function of the inter-fragment distance.
For instance, the excitonic model Hamiltonian (upper triangle) for a dimer at 5~\AA\ distance is
\begin{equation}
\label{e:Hpp_pcm}
{\bf H}^\mathrm{PCM}_\mathrm{pen-pen} =\ \ 
\begin{NiceArray}{rrrrl}
\mathrm{FE}_1 & \mathrm{FE}_2 & \mathrm{CT}_{1\to2} & \mathrm{CT}_{2\to1} \\
1.897 & 0.018 &   0.056 &   -0.046 &  \mathrm{FE}_1  \\
      & 1.897 &  -0.045 &    0.056 &  \mathrm{FE}_2  \\
      &       &  2.508  &    0.000 &  \mathrm{CT}_{1\to2} \\
      &       &         &    2.508 &  \mathrm{CT}_{2\to1} \\
\CodeAfter 
  \SubMatrix({2-1}{5-4})      
\end{NiceArray}
\end{equation}

Figure~\ref{fig:pentacene_pcm} follows the same presentation scheme used for gas-phase results in Figure~\ref{fig:pentacene_gas}, with many similarities between the two data sets.
In the following, we recall the main results and focus on the differences introduced by the polarizable embedding.
The energies of diabatic states are shown in Figure~\ref{fig:pentacene_pcm}a. 
The energies of the diabatic FEs is again almost independent of the intermolecular distance and it is little influenced by the PCM environment.  
On the other hand, diabatic CT states are strongly stabilized by the dielectric medium, converging to the gap as $1/(\epsilon_{opt} R)$, the gap also being strongly reduced by screening effects as compared to the gas-phase calculation.  
The effect of the polarizable environment hence brings CT states closer in energy to FEs, thus favoring a stronger hybridization between the two types of excitations.

\begin{figure}[H]
\centering\includegraphics[scale=1.2, clip, trim=0.0cm 4.2cm 0.0cm 0.0cm]{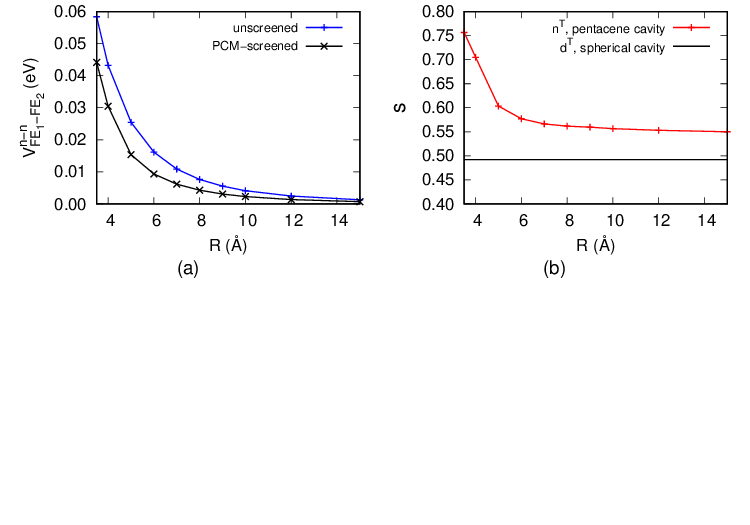}
 \caption{(a) Unscreened and screened (PCM, $\epsilon_\mathrm{opt}=3.5$) Coulomb interaction between FE transition densities.
 Results for the pentacene cofacial dimer as a function of the intermolecular distance $R$. 
 Transition densities were obtained from BSE/$GW$ calculations in PCM.
 (b) Effective screening factor $s$ (Equation~\ref{e:s_eff}) as a function of the intermolecular distance $R$.
 The red points correspond to the ratio between the PCM-screened and unscreened interaction between the transition densities ($n^\mathrm{T}$).  
 The black horizontal line marks the limit expected for point transition dipoles ($\mathbf{d}^\mathrm{T}$)  in spherical cavities at large $R$ (Equation~\ref{e:s_eopt}).} 
    \label{fig:ratio}
\end{figure}

The FE-FE coupling (Figure~\ref{fig:pentacene_pcm}b) obtained with the diabatization scheme closely matches the value computed as screened electrostatic interaction between transition densities (Equation~\ref{e:dens_dens_V_env}), recovering the screened dipolar approximation at large $R$.
\revision{The Coulomb coupling between the transition densities of the two FEs in PCM is significantly reduced as compared to the unscreened interaction, as shown in Figure~\ref{fig:ratio}a.
We recall the importance of computing screened and unscreened interactions between the same transition densities, here corresponding to those obtained from BSE/$GW$ calculations in PCM.
This is crucial for disentangling the effect of dielectric screening from the effect of the polarizable environment on the transition density.
}
Figure~\ref{fig:ratio}b shows the effective screening factor, corresponding to the ratio between screened and unscreened Coulomb interaction between transition densities (see Eq.~\ref{e:s_eff}), as a function of $R$.
The screening factor decays monotonically from 
$\sim$0.76 at 3.5~\AA\ 
distance to reach an asymptotic value of $ \sim$0.55 at large $R$. The trend is similar to what was reported by Mennucci and coworkers for a large set of chromophores in PCM.\cite{scholes2007solvent,curutchet2007solvent}
We further note that the limiting $s$ values is within 12\% from the analytical result for point dipoles in spherical cavities  (Eq.~\ref{e:s_eopt}), i.e.
$s=0.49$ for $\epsilon_\mathrm{opt}=3.5$ (horizontal line in Fig.~\ref{fig:ratio}a). 
Such reasonable agreement shall be considered accidental, since for anisotropic cavities, the screening factor largely depends on the dipole orientation (see SI, Figure~S1).

As for the FE-CT couplings (see Figure~\ref{fig:pentacene_pcm}c,d), we observe an exponential decay of these couplings similar to the gas-phase result.
The quality of the one-electron approximation (PBE0 couplings) is comparable to what obtained in the absence of PCM embedding.
Also in the case of PCM, the direct coupling between CT states is negligible for $R>5$ \AA.

\subsection{Pentacene molecular crystal}
\label{s:crystals}
We now apply our projective diabatization scheme to the pentacene crystal, which allows us to discuss some important and often overlooked aspects of the \emph{ab initio} calculation of the parameters entering model Hamiltonians for extended systems. 
Before proceeding, we recall two technical aspects of our calculations that will be crucial for the quality of the results presented in the following. 
First, all calculations in the crystal employed a QM/MM scheme based on a polarizable atomistic model in the ground-state DFT calculation, where the chosen QM region (single molecule, dimer, etc.) is embedded in the electrostatic potential of the bulk crystal. 
The environmental screening of electronic excitations at the $GW$ and BSE level has been described at the PCM level ($\epsilon_r=3.5$, see Section~\ref{s:environ} for details).
The proper account of the electrostatic landscape and dielectric screening of the environment is essential for the reliable transfer of the parameters calculated from a small QM subsystem to crystals.
In the second instance, the definition of the localized (diabatic) basis functions relies on a single (DFT, $GW$, BSE) calculation for each of the symmetry-unique molecules in the unit cell.
Translational invariance is enforced by creating translational copies of unique fragments and defining the basis functions accordingly. 
This ensures phase consistency among all basis FE and CT states throughout a virtually infinite crystal, hence determining unambiguously the signs of the off-diagonal elements in model Hamiltonians.\cite{Yamagata12,Hetstand2018}

\begin{figure}[H]
\hspace{-3.5cm}\begin{subfigure}{0.30\textwidth}
\includegraphics[scale=0.4, clip, trim= 4.5cm 6.6cm 5.5cm 1.5cm]{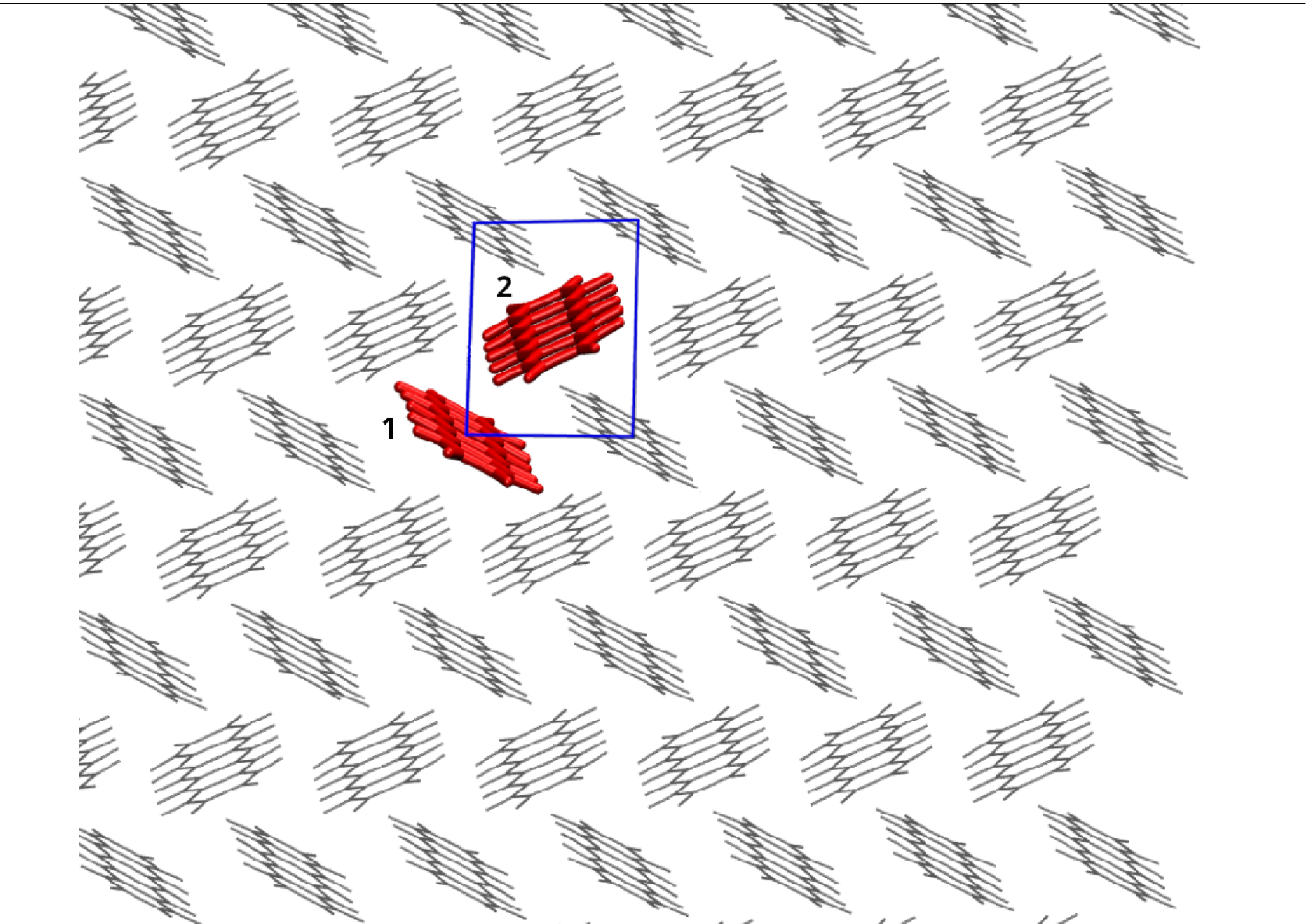}
\caption{}
\end{subfigure}
{\color{white}aaaaaaaaaaaaaaaaaaa}
\begin{subfigure}{0.30\textwidth}
\includegraphics[scale=0.4, clip, trim= 4.5cm 6.6cm 5.5cm 1.5cm]{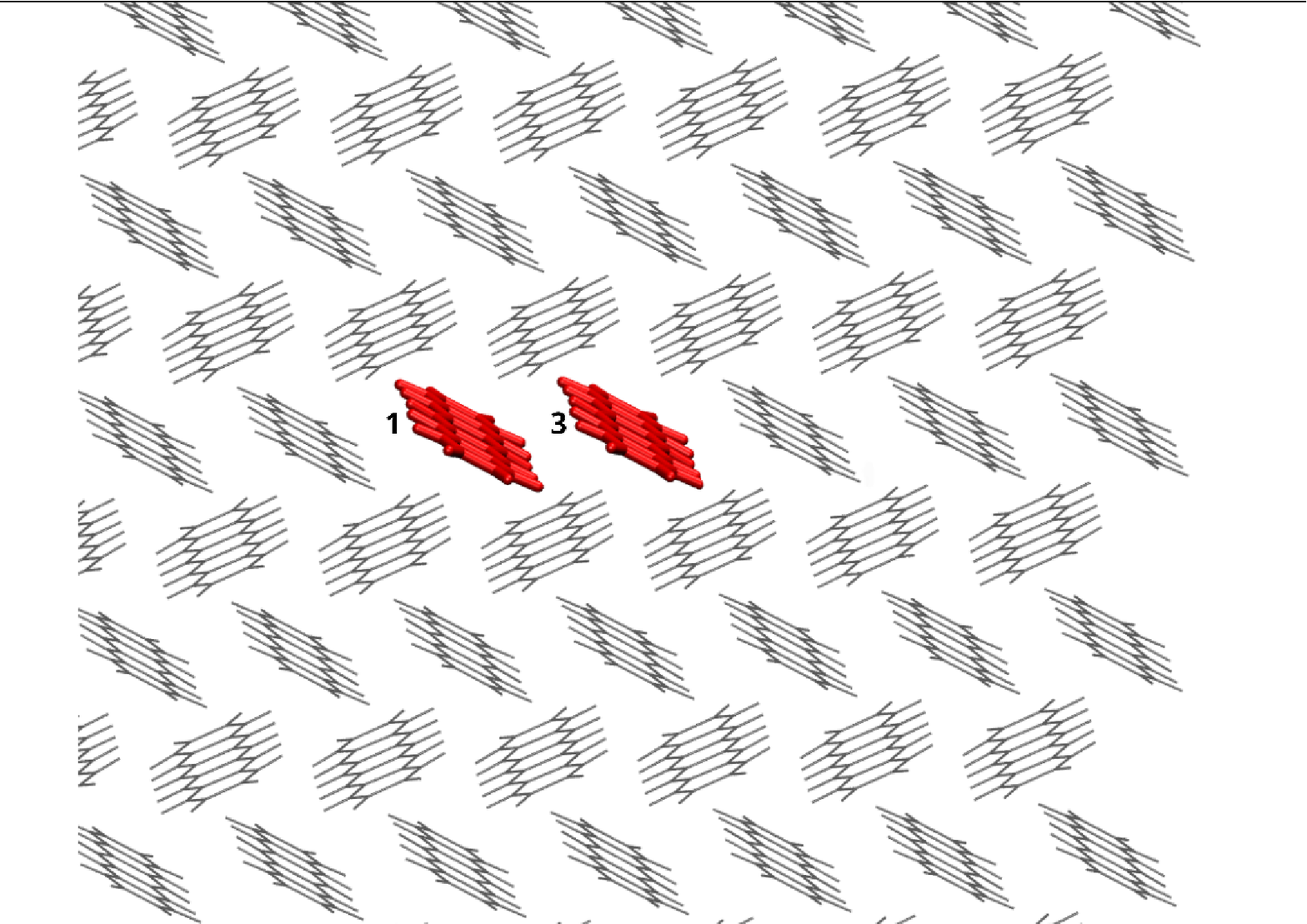}
\caption{}
\end{subfigure} \\
\hspace{-3.5cm}\begin{subfigure}{0.30\textwidth}
\includegraphics[scale=0.4, clip, trim= 4.5cm 6.6cm 5.5cm 1.5cm]{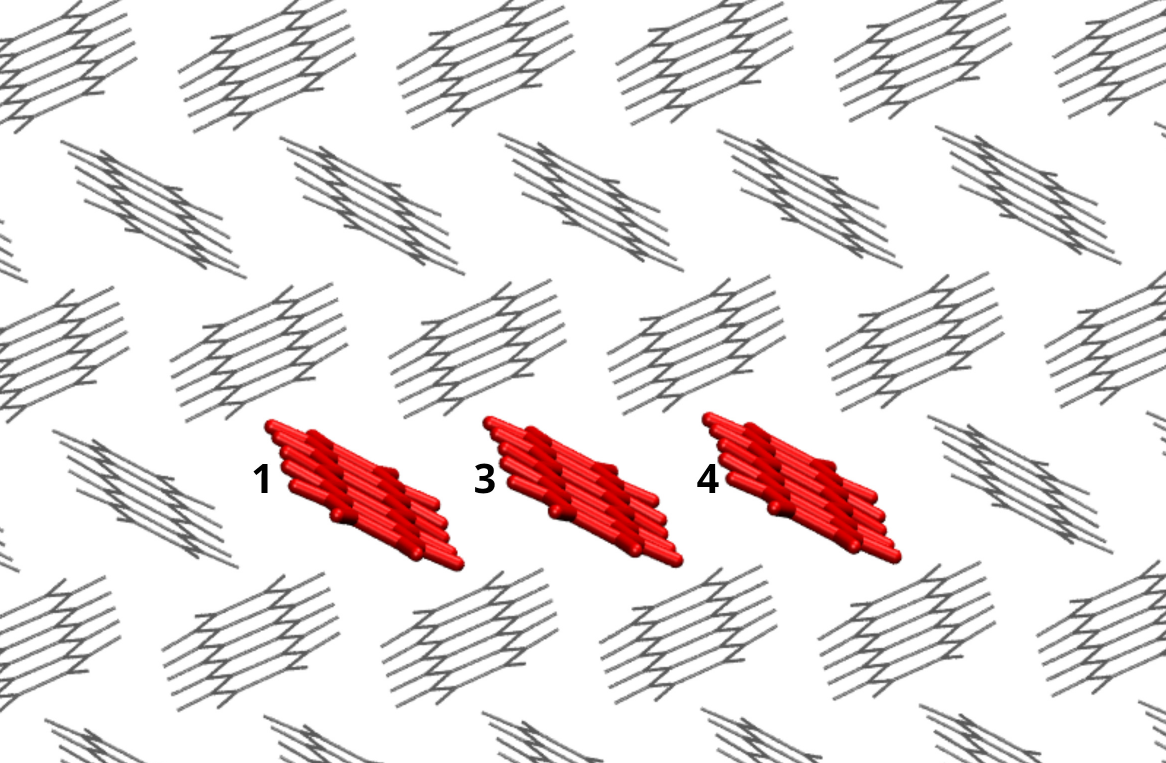}
\caption{}
\end{subfigure}
{\color{white}aaaaaaaaaaaaaaaaaaa}
\begin{subfigure}{0.30\textwidth}
\includegraphics[scale=0.4, clip, trim=4.5cm 6.6cm 5.5cm 1.5cm]{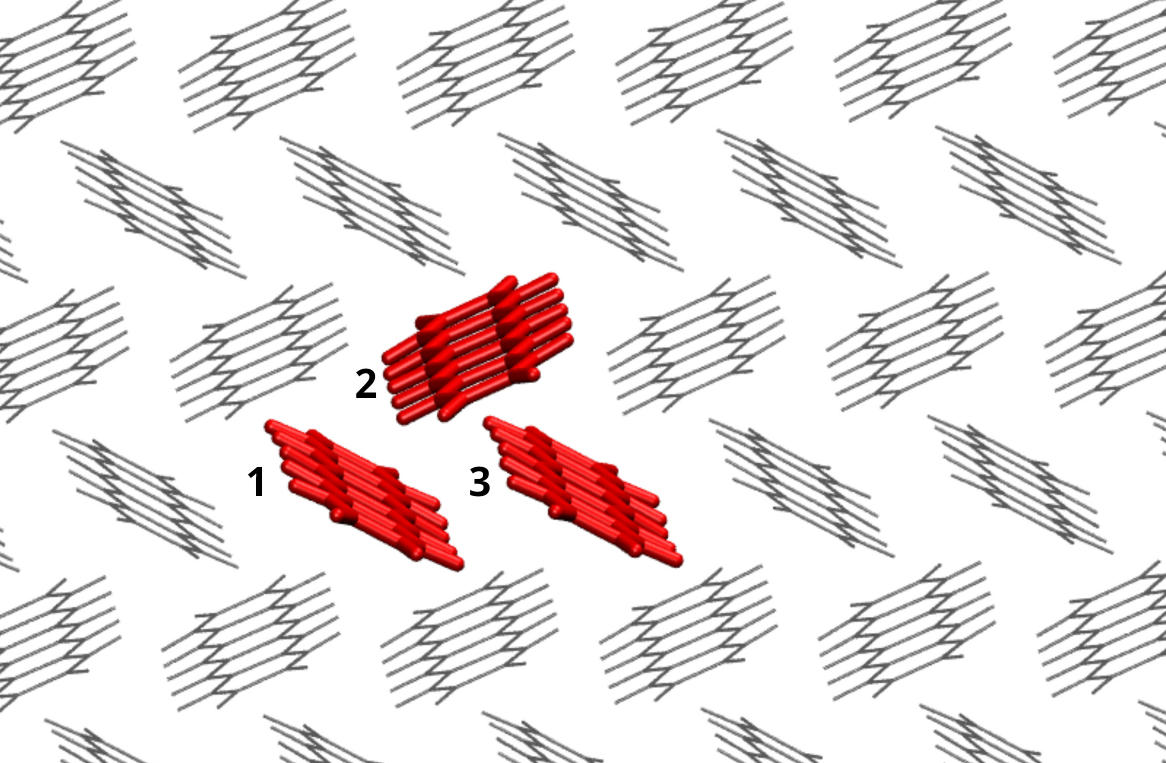}
\caption{}
\end{subfigure}
\caption{Rendering of the pentacene crystal, highlighting in red the (a,b) dimers  and (c,d) trimers employed as QM subsystem in BSE QM/MM calculations.
Molecules drawn in black correspond to the MM embedding environment.
The blue frame in panel (a) displays the crystal unit cell.
The numbering of molecular fragments will be used to label the basis functions used to set up the excitonic model.
}
\label{fig:tetramer}
\end{figure}

The pentacene crystal contains two symmetry-unique molecules in the unit cell arranged in a herringbone pattern (see Figure~\ref{fig:tetramer}).
Two possible choices for molecular dimers to be chosen as QM subsystems are the herringbone  (\{1,2\}, shown in Figure~\ref{fig:tetramer}a) and the parallel (\{1,3\}, Figure~\ref{fig:tetramer}b) dimers.
Consistent with Section~\ref{s:dimers}, the basis for pentacene dimers includes two FEs and two CT states (intermolecular HOMO$\to$LUMO transitions). 
For these two dimers, our projective scheme yields the two effective Hamiltonians:
\begin{equation}
\label{e:H12}
{\bf H}_\mathrm{\{1,2\}} =\ \ 
\begin{NiceArray}{rrrrl}
\mathrm{FE}_1 & \mathrm{FE}_2 & \mathrm{CT}_{1\to2} & \mathrm{CT}_{2\to1} \\
2.013 & -0.005 & -0.057 &   0.099 & \mathrm{FE}_1 \\
      &  1.906 &  0.104 &  -0.058 & \mathrm{FE}_2 \\
      &        &  2.414 &  -0.003 & \mathrm{CT}_{1\to2} \\
      &        &        &   2.295 & \mathrm{CT}_{2\to1} \\
\CodeAfter 
  \SubMatrix({2-1}{5-4})      
\end{NiceArray}
\end{equation}
and
\begin{equation}
\label{e:H13}
{\bf H}_\mathrm{\{1,3\}} =\ \ 
\begin{NiceArray}{rrrrl}
\mathrm{FE}_1 & \mathrm{FE}_3 & \mathrm{CT}_{1\to3} & \mathrm{CT}_{3\to1} \\
2.002 & -0.012 & -0.026 & -0.048 & \mathrm{FE}_1 \\
      & 2.002  & -0.048 & -0.026 & \mathrm{FE}_3 \\
      &        &  2.506 &  0.001 & \mathrm{CT}_{1\to3} \\
      &        &        &  2.508 & \mathrm{CT}_{3\to1} \\
\CodeAfter 
  \SubMatrix({2-1}{5-4})      
\end{NiceArray}
\end{equation}


It would be tempting to use these Hamiltonian matrix elements obtained from these and other nearest-neighbor dimers to build the crystal Hamiltonian. 
However, some CT-CT couplings corresponding to charge hopping between nearest-neighbor fragments (i.e. those where the geminate charge sits on a third molecule, labeled HT and ET in Figure~\ref{f:sketch}) cannot be obtained from dimer calculations. 
For instance, the direct coupling between CT$_{1\rightarrow 2}$ and CT$_{1\rightarrow 3}$ is not accessible from the two previous dimer calculations. 
This is an ET process that can be approximated as the LUMO-LUMO interaction among the fragment's Kohn-Sham orbitals (i.e. $\langle \phi^2 _{L} |\widehat{\textrm{H}}_{KS}|\phi^3 _L \rangle$), for which we would expect a magnitude comparable to FE-CT couplings.

Trimer calculations have been thus performed to include these effects, specifically computing the $\{1,2,3\}$ and $\{1,3,4\}$ systems as defined in Figure~\ref{fig:tetramer}c,d.  
The construction of the diabatic basis for trimers follows the same prescription adopted for dimers, including 3 FEs and 6 CT states, resulting in a nine-state model. 
\revision{The upper triangular part of the Hamiltonians (in eV) for the two trimers reads:}
\begin{equation}
\label{e:H123}
{\bf H}_\mathrm{\{1,2,3\}} =\ \ 
\begin{NiceArray}{rrrrrrrrrl}
\mathrm{FE}_1 & \mathrm{FE}_2 & \mathrm{FE}_3 & \mathrm{CT}_{1\to2} & \mathrm{CT}_{2\to1} & \mathrm{CT}_{2\to3} & \mathrm{CT}_{3\to2} & \mathrm{CT}_{1\to3} & \mathrm{CT}_{3\to1}\\
2.031 &   -0.003&   -0.013 & -0.059& 0.086&   0.002& -0.002&   -0.024& -0.048& \mathrm{FE}_1 \\
&    1.919&   -0.003 &  0.084&-0.056&   0.068& -0.139&   -0.003& -0.003& \mathrm{FE}_2 \\
&   &    2.005 & -0.002& 0.000&  -0.138&  0.061&   -0.048& -0.025& \mathrm{FE}_3 \\
&   &   &  2.432&-0.002&  -0.002& -0.048&    0.056&  0.000&\mathrm{CT}_{1\to2} \\
&   &   & & 2.332&  -0.024& -0.003&    0.000& -0.142&\mathrm{CT}_{2\to1} \\
&   &   & & &   2.340& -0.003&    0.087&  0.000&\mathrm{CT}_{2\to3} \\
&   &   & & &  &  2.372&   -0.001& -0.053&\mathrm{CT}_{3\to2} \\
&   &   & & &  & &    2.553&  0.000&\mathrm{CT}_{1\to3} \\
&   &   & & &  & &   &  2.501&\mathrm{CT}_{3\to1} \\
\CodeAfter 
  \SubMatrix({2-1}{10-9})      
\end{NiceArray}
\end{equation}

\begin{equation}
\label{e:H134}
{\bf H}_\mathrm{\{1,3,4\}} =\ \ 
\begin{NiceArray}{rrrrrrrrrl}
\mathrm{FE}_1 & \mathrm{FE}_3 & \mathrm{FE}_4 & \mathrm{CT}_{1\to3} & \mathrm{CT}_{3\to1} & \mathrm{CT}_{3\to4} & \mathrm{CT}_{4\to3} & \mathrm{CT}_{1\to4} & \mathrm{CT}_{4\to1}\\
 2.019&   -0.012&   -0.001 & -0.026&-0.048&  -0.001& -0.001&    0.005& -0.001& \mathrm{FE}_1 \\
&    2.002&   -0.012 & -0.048&-0.026&  -0.026& -0.048&    0.000&  0.000& \mathrm{FE}_3 \\
&   &    2.019 & -0.001&-0.001&  -0.048& -0.026&   -0.001&  0.005& \mathrm{FE}_4 \\
&   &   &  2.526& 0.001&  -0.001&  0.000&   -0.023&  0.000&\mathrm{CT}_{1\to3} \\
&   &   & & 2.514&   0.005& -0.001&    0.000& -0.049&\mathrm{CT}_{3\to1} \\
&   &   & & &   2.512&  0.001&   -0.049&  0.000&\mathrm{CT}_{3\to4} \\
&   &   & & &   &  2.528&    0.000& -0.023&\mathrm{CT}_{4\to3} \\
&   &   & & &   &  &    2.809&  0.000&\mathrm{CT}_{1\to4} \\
&   &   & & &   &  &    &  2.814&\mathrm{CT}_{4\to1} \\
\CodeAfter 
  \SubMatrix({2-1}{10-9})      
\end{NiceArray} .
\end{equation}

The ET and HT couplings, missing in dimer calculations, are sizeable, with a magnitude of tens of meV.
The comparison of the Hamiltonian obtained for the two trimers reveals that  the ET coupling is almost independent of the position of the hole, namely 
$\langle \mathrm{CT}_{2\to1} | \mathrm{H}_{\{123\}}| \mathrm{CT}_{2\to3} \rangle=-24$~meV and $\langle \mathrm{CT}_{4\to1} | \mathrm{H}_{\{234\}}| \mathrm{CT}_{4\to3} \rangle=-23$~meV.
Similarly, also HT is almost unaffected by the electron position, i.e.
$\langle \mathrm{CT}_{1\to2} | \mathrm{H}_{\{123\}}| \mathrm{CT}_{3\to2} \rangle=-48$~meV and
$\langle \mathrm{CT}_{1\to4} | \mathrm{H}_{\{134\}}| \mathrm{CT}_{3\to4} \rangle =-49$~meV.
The fact that these hopping terms barely depend on the position of the geminate charge, indicates that the screened-Coulomb ($W$) and exchange $v$ terms in the BSE Hamiltonian (see Equation~\ref{e:bse_AB}) are almost negligible, as somewhat expected for loosely overlapping orbitals localized on different molecules.
We note that these two HT and ET couplings are similar to their one-body approximation, 
$\langle \phi^1_\mathrm{HOMO} |\widehat{\textrm{H}}_{KS}|\phi^3_\mathrm{HOMO} \rangle=-29$~meV and 
$\langle \phi^1_\mathrm{LUMO} |\widehat{\textrm{H}}_{KS}|\phi^3_\mathrm{LUMO} \rangle=-65$~meV for HT and ET, respectively. However, the quantitative discrepancy between these Kohn-Sham and BSE/$GW$ couplings attests to the importance of many-body effects for these interactions.

By performing the diabatization on different embedded dimers and trimers from the pentacene crystal structure, we recognize the presence of some diagonal and off-diagonal matrix elements that can be obtained from multiple calculations, which are reported in Table~\ref{tab:me_pen_crys}.
The diagonal energy of the FE of fragment 1 is the only parameter that is accessible in all calculations. 
Its variability  is below 30~meV, which corresponds to 1.5\%  of the excitation energy.
A smaller fluctuation among different choices of  the QM region is obtained for the energy of the FE of fragment 3.

The diagonal energies of CT states are the parameters that are mostly sensitive to the environment, as they consist of localized charges (in contrast with FE that are neutral species) that directly probe the local electrostatic potential within the crystal, and that experience the response of the polarizable medium to their strong dipole field. 
The largest variability is found for the energy of CT$_{1\to3}$, i.e. 47~meV,  below 2\% of the excitation energy.
This is a very positive result, being this variability smaller than  the accuracy of the BSE/$GW$-based QM/MM methodology (typically 0.1-0.2~eV).
For comparison, calculation not accounting for any embedding (i.e. computing dimers and trimers in the gas phase), yields CT states whose energy varies in the range of 0.48~eV, one order of magnitude larger than embedded calculations.
We note that most of the spread is cured by the electrostatic embedding in the ground state, which compensates for the otherwise different electrostatic fields at molecular sites developing in gas-phase dimers and trimers.
Polarization effects on excitations further reduce the spread in CT states energies obtained from calculations on different QM subsystems and, most importantly, are crucial to quantitatively reproduce excitation energies in a condensed phase.
The spread in the values obtained for off-diagonal matrix elements is found to be smaller than 1~meV, as shown in Table~\ref{tab:me_pen_crys}.

In general, electrostatic and dielectric embedding are found to be key ingredients for the transferability of the parameters from BSE QM/MM calculations on small embedded molecular clusters to crystals.
A proper embedding ensures the stability of the matrix elements derived from our projective diabatization scheme against the arbitrary choice of the QM subsystem, as is demonstrated here in the paradigmatic case of the pentacene crystal.
This attests to the robustness and internal consistency of our approach.

\begin{table}[H]
 \centering
 \begin{tabular}{l rrrr}
   &  $\{1,2\}$ & $\{1,3\}$ & $\{1,2,3\}$ & $\{1,3,4\}$ \\ \hline
 $\langle \mathrm{FE}_{1} | \mathrm{H} | \mathrm{FE}_{1} \rangle$  & 2.013 & 2.002 & 2.031 & 2.019 \\
 $\langle \mathrm{FE}_{3} | \mathrm{H} | \mathrm{FE}_{3} \rangle$  &      & 2.002 & 2.005 & 2.002 \\
 $\langle \mathrm{CT}_{1\to3} | \mathrm{H} | \mathrm{CT}_{1\to3} \rangle$ &       & 2.506 & 2.553 & 2.526 \\
 $\langle \mathrm{CT}_{3\to1} | \mathrm{H} | \mathrm{CT}_{3\to1} \rangle$ &       & 2.509 & 2.501 & 2.514 \\
 $\langle \mathrm{FE}_{1} | \mathrm{H} | \mathrm{FE}_{3} \rangle$  &      & $-0.012$  & $-0.013$ & $-0.012$ \\   
 $\langle \mathrm{FE}_{1} | \mathrm{H} | \mathrm{CT}_{1\to3} \rangle$ & & $-0.026$ & $-0.024$ & $-0.026$ \\
 $\langle \mathrm{FE}_{1} | \mathrm{H} | \mathrm{CT}_{3\to1} \rangle$ & & $-0.048$ & $-0.048$ & $-0.048$ \\
 $\langle \mathrm{FE}_{3} | \mathrm{H} | \mathrm{CT}_{1\to3} \rangle$ & & $-0.048$ & $-0.048$ & $-0.048$ \\
 $\langle \mathrm{FE}_{3} | \mathrm{H} | \mathrm{CT}_{3\to1} \rangle$ & & $-0.026$ & $-0.025$ & $-0.026$ 
 \end{tabular}
 \caption{Excitonic model matrix elements (in eV) for the pentacene crystal obtained with the projective diabatization scheme based on embedded BSE calculations (QM/MM).
 For a given matrix element, we report the values obtained from calculations performed on different dimers and trimers as QM systems.
 The remarkable stability of these energies as obtained from different calculations testifies to the quality and consistency of the method.  }
    \label{tab:me_pen_crys}
\end{table}

As a final illustration of the quality of the excitonic models that can be built with the present diabatization scheme, we consider the case of a pentacene tetramer embedded in the crystal structure (fragments $\{1,2,3,4\}$ as defined in Figure~\ref{fig:tetramer}).
This represents a fairly large system that can be computed as a whole with embedded BSE calculations.
The corresponding transition energies, shown in the left column of Table~\ref{tab:exc_tetramer}, serve as a reference for the eigenvalues of the excitonic model Hamiltonian.
The latter has been assembled in a modular fashion with matrix elements obtained from calculations on molecular trimers (see SI).
The exciton energies obtained with the model (right column in Table~\ref{tab:exc_tetramer}) are in excellent agreement with reference BSE ones.
This is especially true in the low-energy region the model is designed for, with an absolute difference within 11~meV for the first 8 excitons.
The agreement progressively worsens at higher energies, up to reach 88~meV for the highest-energy state in Table~\ref{tab:exc_tetramer}.
This is consistent with the fact that the assumption of the model breaks down as we move to higher energies, where excitation manifolds become more and more dense and the chosen minimal basis becomes insufficient to describe the corresponding states.
Such an overall excellent result for the embedded pentacene tetramer, together with the successful consistency checks on equivalent matrix elements derived from different calculations, demonstrate the quality of our BSE-based projective diabatization scheme and its potential for application in future studies.

\begin{table}[H]
    \centering
    \begin{tabular}{ccc}
 BSE &    Model Hamiltonian   \\ \hline
   1.842  &   1.844   \\
   1.943  &   1.954   \\
   2.003  &   1.993   \\
   2.021  &   2.018   \\
   2.248  &   2.252   \\
   2.370  &   2.371   \\
   2.382  &   2.386   \\
   2.415  &   2.414   \\
   2.460  &   2.510   \\
   2.507  &   2.531   \\
   2.587  &   2.589   \\
   2.660  &   2.653   \\
   2.680  &    n.a.       \\
   2.698  &    n.a.       \\
   2.871  &   2.822   \\
   2.915  &   2.827   \\ \hline
 \end{tabular}
 \caption{Comparison between the BSE excitation energies (in eV) for the pentacene $\{1,2,3,4\}$ tetramer embedded in the crystal, and those obtained with the model Hamiltonian. 
 The latter does not include the states CT$_{2\to4}$ and CT$_{4\to2}$ in the basis so two excitons are not available (n.a.) within the model. }
 \label{tab:exc_tetramer}
\end{table}

\clearpage
\section{Conclusions}
\label{s:conclusions}

In this work, we have presented a general and versatile approach to building effective excitonic models from \emph{ab~initio} many-body calculations, namely the $GW$ and the BSE Green's function formalisms.
The derived excitonic models inherit the ability of BSE/$GW$ to accurately describe FE and CT excitations, thanks to the explicit account of non-local electronic correlations and especially the excitonic e-h interaction.
The combination of BSE/$GW$ formalisms with an effective description of the embedding environment in a QM/MM framework, grants access to excitonic models for condensed-phase systems. 
The merits of our BSE-based diabatization scheme have been illustrated and validated through specific examples of model molecular dimers and a molecular crystal, taking pentacene as a paradigmatic test case.

The proposed multi-state projective diabatization scheme starts from the definition of basis functions (FEs localized on single fragments and inter-fragment CT states) and derives the corresponding excitonic Hamiltonian in order to match target exciton energies and wavefunctions from BSE. 
It is worth recalling some important distinctive aspects of the methodology:
(i) It preserves the non-orthogonality of the basis (usually neglected in excitonic models) that is an essential requisite for upscaling the model to larger systems;
(ii) It naturally ensures consistent phase relationships between the excitonic basis on translationally-equivalent molecules, which is necessary for the correct description of the photophysical properties of molecular crystals (e.g. exciton dispersion, quantum dynamics);
(iii) It accounts for electrostatic and dielectric embedding, leading to model parameters that are explicitly calculated for a condensed phase and that minimally depend on the arbitrary choice of the QM subsystem (QM/MM partitioning).
The combination of these features allows one to derive exciton energies and couplings in the diabatic representation that are computed for relatively small QM systems (molecular dimers or trimers) and that can be readily transferred to model Hamiltonians for extended systems,  achieving an accuracy comparable to the original BSE calculation.

The proposed projective diabatization scheme paves the way for an accurate, insightful, and  computationally-efficient description of the excited states of molecular solids based on many-body \emph{ab~initio} theories, setting solid grounds for future studies on complex systems of timely interest.

\clearpage
\begin{acknowledgement}
G.D. acknowledges the contribution of Jing Li in the first steps of this research and thanks Samuele Giannini for stimulating discussions.
This work received financial support from the French ``Agence Nationale de la Recherche", project RAPTORS 
(ANR-21-CE24-0004-01).
This work was performed using HPC resources from GENCI-TGCC (Grant No. 2021-A0110910016).
\end{acknowledgement}

\begin{suppinfo}
Explicit expression for a Frenkel-CT excitonic model Hamiltonian. 
Supplementary calculation results.
Derivation of Eq.~\ref{e:s_eopt} for the screened interaction between point dipoles in spherical cavities and additional results for screened dipolar interaction in realistic molecular cavities.
Bartels-Stewart algorithm for the solution of the Sylvester’s equation.
\end{suppinfo}

\bibliography{xcBSE}
 
\newpage
\section*{TOC Graphics}
\begin{center}
\includegraphics[scale=0.5]{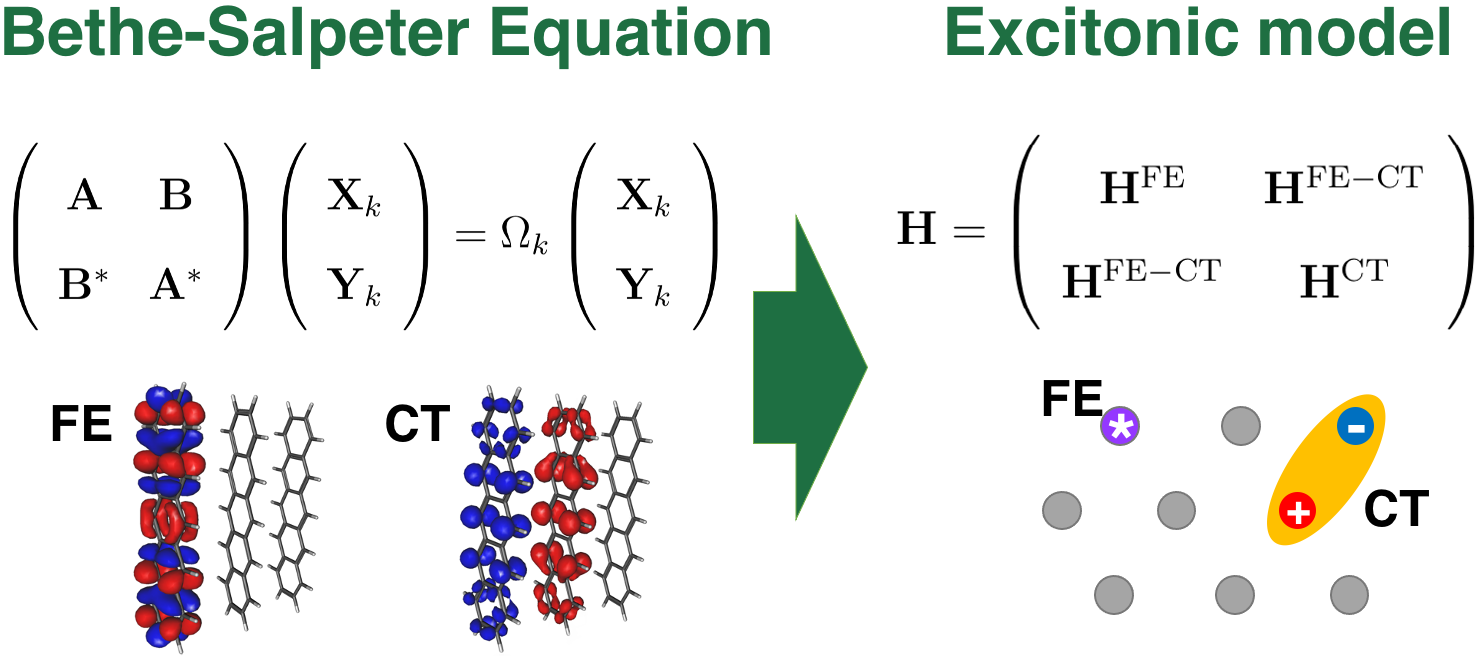}
\end{center}

\end{document}